\documentclass[prd,aps,epsf,amsfonts,floats,amssymb,amsmath,nofootinbib]{revtex4}

\usepackage{color}

\newcommand{\be}{\begin{equation}}\newcommand{\ee}{\end{equation}}
\newcommand{\bea}{\begin{eqnarray}}\newcommand{\eea}{\end{eqnarray}}
\newcommand{\beaa}{\begin{eqnarray}}\newcommand{\eeaa}{\end{eqnarray}}
\newcommand{\ba}{\begin{array}}\newcommand{\ea}{\end{array}}
\newcommand{\bit}{\begin{itemize}}\newcommand{\eit}{\end{itemize}}
\newcommand{\ben}{\begin{enumerate}}\newcommand{\een}{\end{enumerate}}

\definecolor{darkred}{rgb}{.8,0,0}

\def\lf{\left}

\def\ri{\right}

\def\al{\alpha}
\def\de{\delta}

\def\la{\lambda}

\def\om{\omega}

\def\1{{_{1}}}\def\2{{_{2}}}
\def\ZzZ{{\hbox{\tenrm Z\kern-.31em{Z}}}}
\def\CcC{{\hbox{\tenrm C\kern-.45em{\vrule height.67em width0.08em depth-
.04em \hskip.45em }}}}

\newcommand{\lab}{\label}
\newcommand{\non}{\nonumber}

\newcommand{\bc}{\begin{center}}
\newcommand{\ec}{\end{center}}

\begin{document}
%
\title{Gauge theory and two level systems}
\author{A. Bruno${}^{\dag}$,  A. Capolupo${}^{\flat}$, S. Kak${}^{\natural}$, G. Raimondo${}^{\dag}$
and G. Vitiello${}^{\flat}$\footnote{corresponding author: vitiello@sa.infn.it
}}
\address{
${}^{\dag}$ Dipartimento di Fisica ""E.R.Caianiello"" and INFN,
Universit\`a di Salerno, I-84100 Salerno, Italy, \\
${}^{\flat}$Dipartimento di Matematica e Informatica, Universit\`a di Salerno
and\\ INFN, Gruppo Collegato Salerno,  I-84100 Salerno, Italy}

\address{${}^{\natural}$Department of Computer Science,
Oklahoma State University,
Stillwater, OK 74078, USA}

\bigskip

\begin{abstract}
We consider the time evolution of a two level system (a two level atom or a qubit) and show that it is characterized by a local (in time) gauge invariant evolution equation. The covariant derivative operator is constructed and related to the free energy. We show that the gauge invariant characterization of the time evolution of the two level system is analogous to the birefringence phenomenon in optics. The relation with Berry-like and Anandan--Aharonov phase is pointed out. Finally, we discuss entropy, environment effects and the distance in projective Hilbert space between two level  states in their evolution.
\end{abstract}

\maketitle

\section{Introduction}

The role played by gauge fields in quantum theories is of crucial physical relevance, to the point that the gauge theory conceptual and formal scheme has become  the gauge theory paradigm in quantum field theory (QFT) of condensed matter physics and high energy physics. Quantum fluctuations characterize the dynamics of quantum systems. Due to quantum fluctuations, quantum systems live on a variety of microscopic configurations. However, a remarkable stability is often observed at the mesoscopic and macroscopic scale. The well known QFT strategy to face the problem of the emergence of mesoscopic and macroscopic stability out of the fluctuating configurations of the system at the microscopic level consists in requiring the local gauge invariance of the system lagrangian. In practice, one requires that terms proportional to $\partial_{\mu}\theta({\bf
x},t)$, arising  from the kinetic term in the Lagrangian when the elementary component
field $\phi({\bf x},t)$ undergoes the local phase transformation $\phi({\bf x},t) \rightarrow \phi'({\bf x},t) =
\exp(ig\theta({\bf x},t))\phi({\bf x},t)$, must be compensated by the transformation of the gauge field
$A_{\mu}({\bf x},t) \rightarrow A_{\mu}'({\bf x},t) -
\partial_{\mu}\theta({\bf x},t)$. In such a sense, the gauge field thus behaves as a `reservoir' for the $\phi$ field system \cite{Celeghini:1992a,Celeghini:1993a,DelGiudice:2006a}. This is a well known story, indeed.

The problem of the stability of macroscopic complex systems arising from fluctuating quantum components is of special interest when one considers, for example, the process of defect formation in the non-equilibrium symmetry breaking phase transitions characterized by an order parameter \cite{Bunkov} (e.g. vortices in superconductors and superfluids, magnetic domain walls in
ferromagnets,  monopoles and cosmic strings in high energy physics and cosmology \cite{volovik1,kib,kib2,zurek1,Alfinito:2001aa,Alfinito:2001mm,difettibook:2011a}, the onset of phase locking among the electromagnetic modes and the matter components in the formation of coherent domains \cite{DelGiudice:2006a}, etc.). In these cases, the gauge structure of the theory reveals itself to be essential and the topological characterization of the macroscopically behaving defect appears to be a global feature arising from the quantum fluctuating system components.

In this paper we show that the gauge theory paradigm, in conjunction with geometric phase properties, may be applied also to the time evolution of a two level system such as a two level atom or a qubit, thus being relevant also in quantum optics and quantum computation studies. As a matter of fact, it has already been realized \cite{Zanardi:1999} that the geometric (Berry-like) phase and the related (gauge field) connection \cite{Wilczek:1984} play a relevant role in quantum computing. Although such a role of the gauge field has already been  recognized, our aim in this paper is to provide, through the explicit construction of the covariant derivative, for simplicity, in the case of a two level system or qubit, the physical implications of such a gauge structure by showing how it is
related to thermodynamical operators such as the free energy operator, and how it provides the analogy of the two level system with the birefringence phenomenon in optics. The novel picture of time evolution of the two level system thus emerges from our discussion: the two level system appears to be embedded in a gauge field background and it evolves in time in such a way that invariance under local in time gauge transformations is preserved  and it simulates the propagation through a birefringent medium. Finally, we also compute the static and dynamic entropy and the distance between the two level system or qubit states in the Hilbert space in terms of the Fubini--Study metric. 

\section{Time evolution and the gauge structure}

In our analysis  we consider for simplicity the familiar example of a two level system, e.g. a two level atom or a qubit or any other system which might be described by the standard orthonormal basis of two unit (pure state) vectors  $|0\rangle$ and $|1\rangle$, $\langle i| j \rangle = \de_{ij}$, $~i,j = 0,1$. They may be thought as being eigenstates of the operator
\be \lab{h} H = \omega_1 |0 \rangle \langle 0| + \omega_2 |1 \rangle \langle 1| ~,
\ee
with eigenvalues $\om_1$ and $\om_2$, respectively: $H |0 \rangle = \om_1 |0 \rangle$ and $H |1 \rangle = \om_2 |1 \rangle$. We assume that $|0\rangle$ and $|1\rangle$ are non degenerate eigenstates of $H$, i.e. that $\om_1 \neq \om_2$. In other words, $\om_1$ and $\om_2$  denote the (two different) values of the quantum number (energy, or charge, or spin, etc.) characterizing  the states $|0\rangle$ and $|1\rangle$, respectively. In the following we will think of them as the energy (the frequencies, in natural units $h = 1 = c$) eigenvalues.

By a convenient rotation in the plane $\{ |0\rangle,|1\rangle \}$, one may then prepare, at some initial time $t_0 = 0$, the superposition of states
\begin{eqnarray}\lab{phi}
|\phi  \rangle & = &
\alpha  \;|0\rangle \;+\;  \beta  \; |1\rangle \,,~
\\ \lab{psi}
|\psi  \rangle & = &
-\beta  \;|0\rangle \;+\; \alpha  \; |1\rangle \,.
\end{eqnarray}
As usual, orthonormality requires that the coefficients $\alpha$ and $\beta$ satisfy the relations $|\alpha |^{2} + |\beta |^{2} = 1$ and $\alpha^{*}\beta-\alpha\beta^{*}=0$. Thus we may set in full generality $\alpha  = e^{i\gamma_{1} } \cos \theta$ and $\beta = e^{i\gamma_{2} } \sin \theta $, with $\gamma_{1}=\gamma_{2}+ n\pi$, $n=0,1,2...$.
Notice that $|\phi  \rangle$ and $|\psi  \rangle$ are not eigenstates of $H$ due to the fact that $\om_1 \neq \om_2$.

In general, in the preparation process we have a limited control (or even no control) on the fluctuations of the $\alpha $ and $\beta $ coefficients (the initialization problem \cite{SK03,SK04,SK05}). However, in some cases, such as in nuclear magnetic resonance and electron spin resonance systems, a good precision may be reached in the control of the initialization problem \cite{NAT2009}.

At time $t$ we have:
\bea
\lab{tev}
|\phi (t)\rangle &=& e^{-iHt} |\phi (0)\rangle  = e^{-i\omega_1 t} (\cos \theta |0\rangle + e^{-i(\omega_2-\omega_1)t} \sin \theta |1\rangle) ~,\\
\lab{t02}
|\psi (t)\rangle &=& e^{-iHt} |\psi (0)\rangle = e^{-i\omega_1 t} ( -\sin \theta |0\rangle \,+\, e^{-i(\omega_2-\omega_1)t} \cos \theta |1\rangle)\,,
\eea
where we have used the notation $|\phi (0)\rangle \equiv |\phi \rangle$ and  $|\psi (0)\rangle \equiv |\psi \rangle$, and, for simplicity, we have considered real $\al$ and $\beta$ ($\gamma_1 = 0 = \gamma_2$). Of course, time evolution preserves the orthonormalization of the states $|\phi \rangle$ and  $|\psi \rangle$ at any time $t$.

By inverting Eqs.~(\ref{tev}) and (\ref{t02}), we obtain the expression for $H$ at any $t$:
\begin{eqnarray} \label{htr}
H = \om_{\phi \phi} |\phi (t) \rangle \langle \phi (t)| + \om_{\psi  \psi}|\psi (t) \rangle \langle \psi (t)| + \om_{\phi \psi} (|\phi (t) \rangle \langle \psi (t)| + |\psi (t) \rangle \langle \phi (t)|) ~,
\end{eqnarray}
where $\om_{\phi \phi}$, $\om_{\psi \psi}$ and $\om_{\phi \psi}$ are given by the time-independent expectation values
\begin{eqnarray}\label{omee}
&& \omega_{\phi \phi} \, = \, \omega_{1}\;\cos^2\theta + \omega_{2}\;\sin^2\theta \,= \, \langle \phi(t)|\; i\partial_t\;|\phi(t)\rangle  ~,\\ [0.2cm]
\label{ommm}
&& \omega_{\psi \psi}\, = \, \omega_{1}\;\sin^2\theta + \omega_{2}\;\cos^2\theta \,=\,\langle \psi(t)|\; i\partial_t\;|\psi(t)\rangle ~,
\\ \label{omem}
&&\omega_{\phi \psi}\, = \, \frac{1}{2}
( \omega_{2}- \omega_{1})\;\sin 2\theta \,= \, \langle \psi(t)|\; i\partial_t\;|\phi(t)\rangle  ~ ,
\end{eqnarray}
and  $\omega_{\phi \psi}=\omega_{\psi \phi}$. Note that  the $\om_{\phi \psi}$ ``mixed term'',  responsible for  ``oscillations'' between the states $|\phi (t)\rangle$ and $|\psi (t) \rangle$, appears in $H$ since $\omega_2 - \omega_1 \neq 0$ and the "mixing angle" $\theta$ is non-vanishing\footnote{Same situation occurs in the mixing of neutrinos and in general of particles with different masses \cite{BHV99}.}.

We thus have $H |\phi(t)\rangle = \,\om_{\phi\phi}\,|\phi (t)\rangle\,+ \,\om_{\phi \psi}\,\,|\psi(t)\rangle$ and $H |\psi(t)\rangle = \,\om_{\psi\psi}\,|\psi (t)\rangle\,+ \,\om_{\phi \psi}\,\,|\phi(t)\rangle$. On the other hand, since $H |\phi(t)\rangle = i\,\partial_{t}\,|\phi(t)\rangle$ and $H |\psi(t)\rangle = i\,\partial_{t}\,|\psi(t)\rangle$ (cf. Eqs.~(\ref{tev}), (\ref{t02}) and (\ref{h})), we get the evolution equations
\begin{eqnarray}\label{HeisenbergComp}
 i\,\partial_{t}\, |\zeta(t)\rangle\,=\,\om_{d}\,|\zeta(t)\rangle\,+\,\om_{\phi \psi}\,\sigma_{1}\,|\zeta(t)\rangle \,,
\end{eqnarray}
where $|\zeta(t)\rangle$ denotes the vector doublet  $|\zeta(t)\rangle\,=\,(|\phi(t)\rangle \,,|\psi(t)\rangle )^{T}$,  $\om_{d}=diag(\om_{\phi\phi },\om_{\psi\psi })$ and $\sigma_{1}$ denotes the Pauli matrix.
By using the notation  $g \,\equiv\,\tan 2\theta \,=\, \frac{2\omega_{\phi \psi}}{\delta \omega}$, with $\delta \omega \equiv \omega_{\psi\psi }-\omega_{\phi\phi }$, we have $\om_{\phi \psi}\,=\,\frac{1}{2} g\, \delta \omega$. We also put  $A_{0} = A_{0}^{(1)}\,\sigma_{1} \,= \,\frac{1}{2} \, \delta \omega \,\sigma_{1}$. Then we may write
\bea\label{covarDer}
D_{t}\,=\,\partial_{t}\,+\,i\,\om_{\phi \psi}\,\sigma_{1}
\,=\partial_{t}\,+\,i \,g\, A_{0}^{(1)}\,\sigma_{1}\,,
\eea
which acts as the covariant derivative, where $g$ and $A_{0}^{(1)}$ play the role of the coupling constant and the (non-abelian) gauge field, respectively (a  similar situation occurs in the different context of neutrino mixing, see ref. \cite{dimauro2010}). The motion equations
(\ref{HeisenbergComp}) now
can be written as
\bea \label{D}
iD_{t}\,|\zeta(t)\rangle\,=\,\om_{d}\,|\zeta(t)\rangle\, .
\eea
It is easy to show that
\bea
iD_{t}'\,|\zeta'(t)\rangle\,=\,\om_{d}\,|\zeta'(t)\rangle\,,
\eea
with
\bea \label{D1}
&{}& D_{t}'\,=\,\partial_{t}\,+\,i \,g\, (A_{0}^{(1)}\,\sigma_{1}\, + \partial_{t}\,\lambda(t)\,\sigma_{1}),\\ [2mm]
&{}& |\zeta'(t)\rangle = e^{-i g \,\lambda(t)\,\sigma_{1} }|\zeta(t)\rangle\,,
\eea
so that, defining $U(t) \,\equiv\, e^{-i g \,\lambda(t)\,\sigma_{1} }$, it is 
\bea
U(t)\,(iD_{t}\,|\zeta(t)\rangle)\,=\,iD_{t}'\,U(t)\,|\zeta(t)\rangle\,
\eea
and
\bea
g \, {A_{0}^{(1)}}' \,\sigma_{1}\, \, =\, U(t)\,g \,A_{0}^{(1)}\,\sigma_{1}\,U^{-1}(t) \,+\, i\, (\partial_{t}\,U(t))\,U^{-1}(t)\, ,
\eea
as it should be indeed for a gauge field transformation (see Eq.~(\ref{D1})).

We can express the above result by saying that the  time evolution of the vector doublet $|\zeta(t)\rangle$ (our two level system or qubit) is controlled by its coupling with a non-abelian gauge field background so to preserve the invariance of the dynamics against local in time gauge  transformations (phase fluctuations).

We also note that since the only non-vanishing component of $A_{\mu}$ is $A_{0}$ and this is a constant ($A_{0}\,\equiv\,\frac{1}{2} \, \delta \omega \,\sigma_{1}$), the field strength  $F_{\mu \nu}$ is identically zero. This is a feature which, for example, occurs in the case where the gauge potential is a pure gauge (with non-singular gauge functions).

\subsection{The gauge field background as a birefringence medium}

We now show that the time evolution described above
can be interpreted in terms of a birefringence phenomenon\footnote{The analogy with birefringence has been considered for the case of neutrino mixing \cite{Weinheimer:2010ar}}.


Contrarily to what we have assumed above, let us assume now that the states $|0\rangle$ and $|1\rangle$ are degenerate states, namely their time evolution ''in the vacuum" is given by
\be \left(\begin{array}{c}|0(t)\rangle \\
 [2mm]
 |1(t)\rangle \\ \end{array}\right)=\left(\begin{array}{cc}e^{-i\omega t}  & 0 \\
 [2mm]
 0 & e^{-i\omega t} \\ \end{array} \right)\left(\begin{array}{c}|0(0)\rangle \\ [2mm]
 |1(0)\rangle \\
 \end{array}\right),
\ee
where $\om = 2\, \pi \, \nu$, and the propagation speed in the vacuum is $v_0 = \la \, \nu$. Suppose then that the propagation occurs in a medium presenting different refraction indexes, $n_1$ and $n_2$ for $|0\rangle$ and $|1\rangle$, respectively, i.e. where the propagation over a given path of length ${\ell}$ occurs in different times, $t_1$ and $t_2$ for $|0\rangle$ and $|1\rangle$, respectively:
\be \lab{speed}
t_1 \, = \, \frac{\ell}{v_1} \, = \, \frac{\ell \,n_1}{v_0} = t \, n_1  ~; \qquad
t_2 \, = \, \frac{\ell}{v_2} \, = \, \frac{\ell \,n_2}{v_0} = t \, n_2 ~,
\ee
where $v_1$ and $v_2$ are the propagation speeds in the medium for $|0\rangle$ and $|1\rangle$, respectively, and $t\,=\,\frac{\ell}{v_0}$.
Time evolution is then described by the phase factors $e^{-i\omega \, t_1} =  e^{-i\omega_{1} t}$ and $e^{-i\omega \, t_2}  =  e^{-i\omega_{2} t}$ for the two states, respectively, where $\omega \, t_{i} \, = \, \om \frac{\ell}{v_0}\, n_i = 2\,\pi \,\nu \,t \,n_i = 2\,\pi \,\nu_{i}\, t \,= \,\om_{i} \,t$, $i\, = \, 1,2$, has been used, together with    $\la_{i} \,\nu \,= \, v_i$,  $\la_{i}\, \nu_{i}\, = \, v_0$ and $n_i \,= \, \frac{v_0}{v_i} \,= \, \frac{\nu_{i}}{\nu}$.  Thus, for the mixed states $|\phi \rangle$ and $|\psi \rangle$ given by Eqs. (\ref{phi}) and (\ref{psi}) we have
\be \lab{dif}
\left(
     \begin{array}{c}
       |\phi(t)\rangle \\
       [2mm]
       |\psi(t)\rangle \\
     \end{array}
   \right)= e^{-i\omega_{1}t}
   \left(
     \begin{array}{cc}
       \cos\theta & ~e^{-i(\omega_{2}-\omega_{1})t}\sin\theta \\
       [2mm]
       - \sin\theta & ~e^{-i(\omega_{2}-\omega_{1})t}\cos\theta \\
     \end{array}
   \right)\left(
            \begin{array}{c}
              |0\rangle \\
              [2mm]
              |1\rangle \\
            \end{array}
          \right),
\ee
which is the time evolution generated by $H$ given by Eq. (\ref{h}) with $\om_{1} \neq \om_{2}$ (cf. Eqs. (\ref{tev}) and (\ref{t02}) ). In conclusion, Eq. (\ref{dif}) shows that, provided that $\om_{1} \neq \om_{2}$, for $\theta \neq \frac{\pi}{4} + \frac{n\,\pi}{2}$, the effect of time evolution through the refractive medium is equivalent to the effect of the background gauge field $A_{0}^{(1)} \, = \,\frac{1}{2}
( \omega_{2}- \omega_{1})\;\cos 2\theta \, = \, \frac{1}{2} \om (n_2 - n_1) \;\cos 2\theta$, which indeed disappears when propagation occurs in the vacuum, $n_1 \, = \,n_2 \, = \,n_0 \, = \,1$ (i.e. $\om_1 \, = \, \om \, = \,\om_2$).


\subsection{The free energy, the topological phases and the Fubini--Study metric}

Let us now write Eqs.~(\ref{HeisenbergComp}) as
\begin{eqnarray}\label{HeisenbergCompH}
 (H\, -\,\om_{\phi \psi}\,\sigma_{1})\,|\zeta(t)\rangle = \,\om_{d}\,|\zeta(t)\rangle\,,
\end{eqnarray}
with the covariant derivative denoted by $H\, -\,\om_{\phi \psi}\,\sigma_{1}$. Then the operator
\begin{eqnarray}\label{FreeEn}
 F \,=\, (H\, -\,\om_{\phi \psi}\,\sigma_{1})
\end{eqnarray}
may be interpreted as the free energy operator, provided that one identifies the term $\om_{\phi \psi}\,\sigma_{1} = g A_{0}$ with the entropy term $T S$ in the traditional free energy expression, where the ``temperature" is $T \, = g$ and the entropy $S = A_{0}$. We thus see that time evolution is controlled  by the free energy (\ref{FreeEn}) where the gauge field plays the role of the entropy. In terms of the states $|\phi(t)\rangle$ and $|\psi(t)\rangle$, the term $T S$ is written as:
\begin{eqnarray} \label{TS}
 T S=  \om_{\phi \psi} (|\phi (t) \rangle \langle \psi (t)| + |\psi (t) \rangle \langle \phi (t)|) ~.
\end{eqnarray}
In order to better understand this feature, it is convenient to consider the geometric phases associated to the system evolution. We observe that it is an easy matter to compute the Berry-like phase.  Indeed, one immediately gets the geometrical phase $\beta_{\phi}$ \cite{AA87}:
\begin{eqnarray}\label{ber1}
\beta_{\phi}&=& \varphi + \int_{0}^{T}
\;\langle \phi(t)|\; i\partial_t\;|\phi(t)\rangle \,dt \;= \; 2 \pi \sin^{2}\theta  \, ,
\end{eqnarray}
which is independent of the  $\omega_i$'s, $i=1,2$, and depends only on the ``mixing angle'' $\theta$. In Eq.~(\ref{ber1}), ${\varphi} \equiv - \frac{2\pi \omega_{1}}{\omega_{2} - \omega_{1}}$, $\omega_1 \neq \omega_2$,  and we exploited the fact that
after a period $T= \frac{2\pi}{\omega_{2} - \omega_{1}}$ it is $|\phi(T)\rangle = e^{i {\varphi}} |\phi(0)\rangle$.

Similarly, for the state $|\psi (t)\rangle$ one finds $\beta_{\psi}\,= 2 \pi \cos^{2}\theta$ and thus $\beta_\psi + \beta_{\phi} = 2\pi$ for any $\theta$.

Another geometric invariant is the Anandan--Aharonov phase discussed in ref.~\cite{AA90}.  It has the advantage to be well defined
also for systems with non-cyclic evolution and is given by $s= 2 \int \Delta \om (t) dt$, where $\Delta \om (t)$ is the variance given by
\bea\label{inv1b}
\Delta \, \om^{2} = \Delta \, \om^{2}_{\phi \phi} = \Delta \, \om^{2}_{\psi \psi} =  \langle \xi(t)|H^2
|\xi(t)\rangle - \langle\xi(t)|H |\xi(t)\rangle^2 = \Delta \omega^2_{\phi \psi}=\omega^2_{\phi \psi}\,,
\eea
with $\xi = \phi,\,\psi$\,.
The relation between the entropy and the geometric invariant $s$ is obtained by considering that
\begin{eqnarray} \label{TSs}
 \int \langle \zeta (t)| T S \sigma_{1}|\zeta (t)\rangle ~dt = \int \langle \zeta (t)| g\, A_{0}^{(1)}|\zeta (t)\rangle ~dt =  2\,\int \om_{\phi \psi} \, dt = s .
\end{eqnarray}
It is interesting to note that the relation between $T S$ and the variance of the energy $\Delta \, \om\,=\,\om_{\phi\psi}$ is through the non-diagonal elements of $H$, namely it is proportional to the energy gap, $\om_2 - \om_1$, between the two levels (cf. Eq.~(\ref{inv1b}) and (\ref{omem})). We also recognize that the integrand $\langle \zeta (t)| g\, A_{0}^{(1)}|\zeta (t)\rangle$ in Eq.~(\ref{TSs}) is related to the {\it adiabatic connection} \cite{Wilczek:1984} emerging in the study of the non-abelian holonomy (generalized Berry phase) \cite{Zanardi:1999}. Moreover, one can also show \cite{BHV99} that these connections are related with the parallel transport of the vectors in the parameter space, as well known \cite{Zanardi:1999}. In the Appendix A we compute the static and the dynamic entropy associated to the reduced density matrix.

Finally, it is instructive to analyze the invariant $s$ in Eq.~(\ref{TSs}) in terms of the distance
between states in the Hilbert space.
 Let us generically denote by  $|\xi (t) \rangle $ either  $|\phi (t) \rangle $ or $|\psi (t) \rangle $. Their evolution is governed by the Schr\"odinger
equation
\bea i\,\,\partial_{t}| \xi  (t) \rangle =\, H |\xi (t) \rangle \,,
\qquad \xi  = \phi, \psi\,. \eea
 Expanding the state
$|\xi(t+ dt) \rangle$ up to the second order in $dt$ and
taking into account that $\frac{d}{dt}H ~=~0$, we have 
\bea\non \,\langle \xi(t)
|\xi(t+ dt) \rangle = 1 - {i dt}
 \,\langle \xi(t) | H |\xi(t) \rangle  - \frac{dt^{2}}{2 }
\,\langle \xi(t) | H^{2}|\xi(t) \rangle +
O(dt^{3})\,,
\\
\eea
and
$
|\langle \xi(t) |\xi (t+ dt) \rangle|^{2} =1-{dt^{2}}
\Delta \omega_{\xi\xi} ^{2}\, + \, O(dt^{3})\,.
$
That is
\bea\label{sigmat} \lf|\langle  \xi (t)
|\xi (t+ dt) \rangle\ri|^{2} = 1-{dt^{2}}
\frac{\omega_{-}^{2}}{4} \sin^{2} 2\theta\, + \, O(dt^{3})\,, \qquad
\xi = \phi, \psi\,,
\eea
where we have used Eqs.(\ref{omee}), (\ref{ommm}) and (\ref{inv1b}) and
\bea
\,\langle\phi(t)|\,
 H ^2\,|\phi(t)\rangle &=&
\omega_{1}^{2}\,\cos^{2}\theta\,
+\, \omega_{2}^{2}\, \sin^{2}\theta\, ,
\\ [2mm]
\,\langle\psi(t)|\,
 H ^2\,|\psi(t)\rangle &=&\omega_{2}^{2}\, \cos^{2}\theta\,
+ \,\omega_{1}^{2}\,\sin^{2}\theta\, .
\eea
We also have
\bea\label{modulusQMem} |\langle \phi(t)
|\psi(t+ dt) \rangle|^{2} = |\langle \psi(t) |\phi(t+
dt) \rangle|^{2} = {dt^{2}} \Delta \omega_{\phi \psi} ^{2}\, +
\, O(dt^{3})\,.
\eea
 The Fubini--Study metric
\cite{AA90} is defined as
 \bea\label{Fubini}
ds^{2}\,=\,2\,g_{\mu\nu}\,dZ^{\mu}\,d\bar{Z}^{\nu}\,=
\,4\,(1\,-\,|\langle  \xi (t) | \xi (t+ dt)
\rangle|^{2})\,,
\eea
 where $Z^{\mu}$ are coordinates in the
projective Hilbert space $\mathcal{P}$, which is the set of rays of the
Hilbert space $\mathcal{H}$. From Eqs.(\ref{sigmat}), (\ref{modulusQMem})
and (\ref{Fubini}), we have the infinitesimal geodetic distance
between the points $\Pi(|\phi(t)\rangle)$ and
$\Pi(|\phi(t+dt)\rangle)$  in the space $\mathcal{P}$
\bea \label{lineFS}
ds\,=\,2\,{\Delta
\omega_{\xi\xi}\,dt}\,=\, {\omega_{-}
\,\sin 2\theta\,  dt}\,.
\eea
The rate of change of this distance is $\frac{ds}{dt} = {\omega_{-}
\,\sin 2\theta} = 2\,{\om_{\phi\psi}}$, with $\om_{-} \equiv \omega_2 - \omega_1 \neq 0$. In the case of the
above two level or qubit states, the Fubini--Study metric coincides with the usual
metric on a sphere of unitary radius:
$ds^{2}\,=\,d\Theta^{2}\,+\,\sin^{2}\Theta\,  d\varphi^{2}$, with
$\Theta =2\,\theta$ ($\theta=$ mixing angle) and $\Theta \in
[0,\pi]$.
Since $\theta$ is constant, we have $ds\,=\,\sin 2 \theta\,  d\varphi$
and, by comparison with Eq.(\ref{lineFS}),
 $ d\varphi\,=\,{\omega_{-} }\,dt\,$. We thus obtain
\bea\label{faseQM} s =
\int \sin 2\theta \, d\varphi\,=\,2\int \om_{\phi\psi} \, dt\, ,
\eea
which is the Anandan--Aharonov invariant (cf. Eq.~(\ref{TSs}) ). Thus,  the Anandan--Aharonov invariant $s$
represents the distance between  evolution states, as
measured by the Fubini--Study metric, in the projective Hilbert space
$\mathcal{P}$.

\section{Conclusions}

In conclusion we have shown that time evolution of a two level system or qubit is controlled by a covariant derivative accounting for the coupling of the state with a (non-abelian) gauge field background so to preserve the invariance of the dynamics against local in time gauge transformations. We have shown that the effect of the gauge field background can be depicted as the effect of a birefringence phenomenon, the gauge field bacground acting as the analogous of the refractive medium.
We have also shown that the covariant derivative plays the role of the free energy with the gauge field acting as the entropy. In such a picture time evolution is controlled  by the free energy. Finally, the relation of our result with the geometric phase and the so-called adiabatic connection has been pointed out. We have also shown that the distance in the projective Hilbert space between two level or qubit evolution states is measured by the Fubini--Study metric in terms of the Anandan--Aharonov geometric invariant. For the reader convenience and for completeness, further comments on the entropy and the environment effects are reported in the Appendix A.

\appendix
\section{Mixed states, entropy and environment effects}

By exploiting the Schmit decomposition theorem (see e.g. \cite{SK01,Par}), one may always "double" the system under study; denote it by ${\cal A}$.  The "doubled" system, denoted by  $\tilde {\cal A}$, is introduced in such a way to work in the composite Hilbert space ${\cal H}_{{\cal A},{\tilde {\cal A}}} \equiv {\cal H}_{\cal A} \otimes {\cal H}_{\tilde {\cal A}}$
with states
$|\Psi_{{\cal A}, \tilde {\cal A}}\rangle  = \sum_n \,\sqrt{w_n} \,|a_n \,{\tilde a}_n \rangle \in {\cal H}_{{\cal A},\tilde {\cal A}}$, $\sum_n w_n = 1$.  The density matrix for mixed states of the system ${\cal A}$, $\rho^{\cal A} = \sum_n w_{n} |a_n\rangle \langle a_n |$, is obtained by tracing out the system $\tilde {\cal A}$:
\be \label{mixroreduc}
\rho^{\cal A} = \sum_n w_{n} |a_n\rangle \langle a_n | =
\sum_{nm} \sqrt{w_{n}w_{m}}\, |a_n  \rangle \langle a_{m}  | \, {\rm Tr}\,(|{\tilde a}_n \rangle \langle {\tilde a}_m |) = {\rm Tr}_{\tilde {\cal A}}\,(\rho^{{\cal A} \oplus \tilde {\cal A}} )
~,
\ee
where the relation $\langle {\tilde a}_m |{\tilde a}_n \rangle  = \delta_{nm}$ has been used.
Then, one can show that the ''tilde"  system $\tilde {\cal A}$ can be interpreted as the thermal bath or reservoir for the original system $\cal A$ \cite{Umezawa:1993yq,Celeghini:1992} and the free energy and the entropy can be
defined\footnote{Such a construction is equivalent to the GNS construction in the $C^*$-algebra formalism and requires the quantum field theory framework \cite{Umezawa:1993yq,Celeghini:1992}.}. The state $|\Psi_{{\cal A}, \tilde {\cal A}}\rangle$ is recognized to be an entangled state of the tilde and non-tilde modes and the entropy provides a measure of the entanglement \cite{dimauro2010,Iorio:2004bt}.

We now compute the static (linear) entropy for the qubit states $|\phi(t)\rangle$ and $|\psi(t)\rangle$ given by Eqs.~(\ref{tev}) and (\ref{t02}), respectively.
One introduces the states $| \tilde 0 \rangle$ and $| \tilde 1 \rangle$ as
 \bea\label{fk1}
|0\rangle & \rightarrow & |0\rangle \otimes|\tilde{0}\rangle\,,
\\
\label{fk2}|1\rangle & \rightarrow & |1\rangle \otimes|\tilde{1}\rangle \,,
\eea
and uses Eqs.~(\ref{fk1}) and (\ref{fk2}) in Eqs.~(\ref{tev}) and (\ref{t02}). The density matrices for the states in ${\cal H}_{{\cal S},{\tilde {\cal S}}}$, where ${\cal S}=\{0,1\}$ and ${\cal \tilde{S}}=\{\tilde{0},\tilde{1}\}$, are denoted by  $\rho_\xi\,=|\xi(t), {\tilde \xi}(t)\rangle\langle\xi(t), {\tilde \xi}(t)|$ where $\xi\,=\,\phi\,,\psi$ and ${\tilde \xi}\,=\,{\tilde \phi}\,,{\tilde \psi} $.
The reduced density matrix  $\rho^{\cal S}_\phi$ (and similarly for $\rho^{\cal S}_\psi$) is obtained by tracing out the tilde-system ${\tilde {\cal S}}$, and vice-versa. Thus one obtains:
\bea
\rho^{\cal S}_\phi &=& Tr_{\cal \tilde{S}}\, \rho_\phi= \cos^2\theta \, |0\rangle\, \langle 0|
+ \sin^2\theta\,|1\rangle \, \langle 1|\,,
\\
\rho^{\cal \tilde{S}}_\phi &=& Tr_{\cal S}\, \rho_\phi= \cos^2\theta \,|\tilde{0}\rangle\, \langle \tilde{0}|\,
+ \sin^2\theta\,|\tilde{1}\rangle \, \langle \tilde{1}|\, .
\eea
The static linear entropies $S_L$ associated to the reduced matrices $\rho^{\cal S}_\phi $ and $\rho^{\cal \tilde{S}}_\phi$
are then:
\bea\label{static-Entropy1}
S_L[\rho^{\cal S}_\phi] &=& 2\,(1-Tr_{\cal S}[(\rho^{\cal S}_\phi)^2]) = \sin^2(2\theta),
\\\label{static-Entropy2}
S_L[\rho^{\cal \tilde{S}}_\phi] &=& 2\,(1-Tr_{\cal \tilde{S}}[(\rho^{\cal \tilde{S}}_\phi)^2]) = \sin^2(2\theta).
\eea

Recall that $\sin^{2} 2\theta = \frac{4}{\om^{2}_{-}}\,\om^{2}_{\phi\psi}$ (cf. Eq.~(\ref{omem})), where $\om_{-} \equiv \omega_2 - \omega_1 \neq 0$.
Let us now compute the dynamic entropy. Consider the state $|\phi(t)\rangle$ in Eq.(\ref{tev}) (we can proceed in a similar way for $|\psi(t)\rangle$) and  express it in terms of the states
 $|\phi(0)\rangle$ and $|\psi(0)\rangle$:
\bea
|\phi(t)\rangle\,=\,A_{\phi \phi}(t)|\phi(0)\rangle \,+\, A_{\phi \psi}(t)|\psi(0)\rangle,
\eea
where $A_{\phi \phi}(t)$ e $A_{\phi \psi}(t)$ are the amplitudes:
\bea
A_{\phi \phi}(t)&=&\langle\phi(0)|\phi(t)\rangle = e^{-i \omega_1 t}\cos^2\theta+e^{-i \omega_2 t}\sin^2\theta, \\
A_{\phi \psi}(t)&=&\langle\psi(0)|\phi(t)\rangle = e^{-i \omega_1 t}\sin\theta\cos\theta+e^{-i \omega_2 t}\sin\theta\cos\theta\,,
\eea
respectively.
The tilde-states $|\tilde{\phi}\rangle$ and  $|\tilde{\psi}\rangle$ are introduced, for any $t$, as :
 \bea\label{fk3}
|\phi\rangle & \rightarrow & |\phi\rangle \otimes|\tilde{\phi}\rangle\,,
\\
\label{fk4}|\psi\rangle & \rightarrow & |\psi\rangle \otimes|\tilde{\psi}\rangle \,.
\eea
The reduced density matrices are now, for any $t$,
\bea
\label{rhoPhi1} \rho^{\cal R}_\phi &=& Tr_{\cal \tilde{R}} \rho_\phi= |A_{\phi \phi}(t)|^2\,|\phi \rangle  \,  \langle \phi|
+\;|A_{\phi \psi}(t)|^2\,|\psi\rangle  \, \langle \psi|,
\\
\label{rhoPhi2}\rho^{{\cal \tilde{R}}}_\phi &=& Tr_{\cal R} \rho_\phi= \;|A_{\phi \phi}(t)|^2\,|\tilde{\phi} \rangle  \,  \langle \tilde{\phi}|
+\;|A_{\phi \psi}(t)|^2\,|\tilde{\psi}\rangle  \, \langle \tilde{\psi}|  ,
\eea
where ${\cal R}=\{\phi,\psi\}$ and ${\cal \tilde{R}}=\{\tilde{\phi},\tilde{\psi}\}$.
The dynamic entropies $S_L$ are
\bea\label{entr1}
S_L(\rho^{\cal R}_\phi)&=&2\,(1-Tr_{{\cal R}}[(\rho^{\cal R}_\phi)^2])= 4\,|A_{\phi \phi}(t)|^2\,|A_{\phi \psi}(t)|^2 = 4\,P_{\phi\rightarrow\phi}(t)\,P_{\phi\rightarrow\psi}(t),
\\\label{entr2}
S_L(\rho^{{\cal \tilde{R}}}_\phi)&=&2\,(1-Tr_{{\cal \tilde{R}}}[(\rho^{{\cal \tilde{R}}}_\phi)^2])= 4\,|A_{\phi \phi}(t)|^2\,|A_{\phi \psi}(t)|^2 = 4\,P_{\phi\rightarrow\phi}(t)\,P_{\phi\rightarrow\psi}(t),
\eea
where $P_{\phi\rightarrow\phi}(t)$ and $P_{\phi\rightarrow\psi}(t)$ are the probabilities
of the transitions $ \phi \rightarrow \phi $ and $ \phi \rightarrow \psi $:
\bea
P_{\phi\rightarrow\psi}(t) &=& \sin^2(2\theta)\sin^2 \lf(\frac{\omega_2-\omega_1}{2}t \ri)\,, \\
P_{\phi\rightarrow\phi}(t) &=& 1-\sin^2(2\theta)\sin^2 \lf(\frac{\omega_2-\omega_1}{2}t \ri)\,,
\eea
respectively.


\begin{thebibliography}{99}

\bibitem{Celeghini:1992a}
E. Celeghini, E. Graziano, K. Nakamura and G. Vitiello, {\it Phys. Lett.} {\bf B 285}, 98 (1992)

\bibitem{Celeghini:1993a}
E. Celeghini, E. Graziano, K. Nakamura and G. Vitiello, {\it Phys. Lett.} {\bf B 304}, 121 (1993)

\bibitem{DelGiudice:2006a}
E. Del Giudice and G. Vitiello, {\it Phys. Rev.} {\bf A 74}, 022105 (2006)

\bibitem{Bunkov} Y. M. Bunkov and H. Godfrin (Eds.),
{\it Topological defects and the non-equilibrium dynamics of
symmetry breaking phase transitions}, NATO Science Series C 549,
(Kluwer Acad. Publ. Dordrecht 2000)


\bibitem{kib} T. W. B. Kibble,
{\it J. Phys.} {\bf A 9},  1387 (1976),
{\it Phys. Rep}
{\bf 67}, 183 (1980) \\
A. Vilenkin,
{\it Phys. Rep.} {\bf 121}, 264 (1985)


\bibitem{kib2} T. W. B. Kibble,
in  {\it Topological defects and the non-equilibrium dynamics of
symmetry breaking phase transitions}, Eds. Y.M. Bunkov and H.
Godfrin, NATO Science Series C 549, (Kluwer Acad. Publ. Dordrecht
2000), p. 7


\bibitem{zurek1} W. H. Zurek,
{\it Phys. Rep.} {\bf 276}, 177 (1997) and refs. therein quoted


\bibitem{volovik1} G. E. Volovik,
in {\em Topological defects and the non-equilibrium dynamics of
symmetry breaking phase transitions}, Eds. Y.M. Bunkov and H.
Godfrin, NATO Science Series C 549, (Kluwer Acad. Publ. Dordrecht
2000), p. 1 and p. 353

\bibitem{Alfinito:2001aa}
  E.~Alfinito and G.~Vitiello,
  {\it Phys.\ Rev.} {\bf B 65}, 054105-5 (2002)


\bibitem{Alfinito:2001mm}
  E.~Alfinito, O.~Romei and G.~Vitiello,
  {\it Mod.\ Phys.\ Lett.} {\bf B 16}, 93 (2002)

\bibitem{difettibook:2011a}
M. Blasone, P. Jizba and G. Vitiello, {\it Quantum field theory and its macroscopic manifestations}, (Imperial College Press, London 2011)

\bibitem{Zanardi:1999} J. Pachos, P. Zanardi and M. Rasetti, {\it Phys. Rev.} {\bf A61},  010305(R) (1999)\\ P. Zanardi and M. Rasetti, Phys. Lett. {\bf A 264}, 94 (1999)

\bibitem{Wilczek:1984} F. Wilczek and A. Zee, {\it Phys. Rev. Lett.} {\bf 52},  2111 (1984)

\bibitem{SK03}
S. Kak, {\it Foundations of Physics} {\bf 29}, 267 (1999) 

\bibitem{SK04}
S. Kak, {\it ACM Ubiquity} {\bf 7 (11)}, 1 (2006) 

\bibitem{SK05}
S. Kak, {\it Information Sciences} {\bf 152}, 195 (2003) 

\bibitem{NAT2009}
M. J. Biercuk, H. Uys, A. P. VanDevender, et al.
{\it Nature} {\bf 458}, 996 (2009) 

\bibitem{BHV99}
M. Blasone, P.A. Henning and G. Vitiello,  {\it Phys.Lett.}
{\bf B 451}, 140 (1999)
\\
  M.~Blasone, A.~Capolupo, E.~Celeghini and G.~Vitiello,
  {\it Phys.\ Lett.}\ {\bf B 674}, 73 (2009)


\bibitem{dimauro2010} M.~Blasone, M.~Di Mauro and G.~Vitiello,
{\it Phys.\ Lett.}\ {\bf B 697}, 238 (2011)

\bibitem{Weinheimer:2010ar}
  C.~Weinheimer,
  {\it Prog.\ Part.\ Nucl.\ Phys.}  {\bf 64 } 205 (2010)  

\bibitem{AA87}
M. V. Berry, {\it Proc.Roy.Soc.London} {\bf A 392}, 45 (1984)

\bibitem{AA90}
J.Anandan and Y.Aharonov {\it Phys. Rev. Lett.} {\bf 65}, 1697 (1990)


\bibitem{SK01}
M. A. Nielsen and I. L. Chuang, {\it Quantum Computation and Quantum Information.} (Cambridge University Press 2000)

\bibitem{Par} G. Auletta, M. Fortunato and G. Parisi, {\it Quantum Mechanics}, Cambridge University Press,  2009

\bibitem{Umezawa:1993yq} H. Umezawa, {\it  Advanced field theory: Micro, macro, and thermal physics}, AIP, N.Y. 1993

\bibitem{Celeghini:1992}
E. Celeghini, M. Rasetti and G. Vitiello, {\it Annals Phys.} {\bf 215}, 156 (1992)

\bibitem{Iorio:2004bt}
A. Iorio, G. Lambiase and G. Vitiello, {\it Annals Phys.} {\bf 309}, 151 (2004)


\end{thebibliography}
\end{document}